\documentclass[preprint,5p,times]{elsarticle}
\usepackage[utf8]{inputenc}
\usepackage{amsmath,amsfonts,amssymb, mathtools, blindtext}
\usepackage{natbib}
\usepackage[skip=0pt]{caption}
\usepackage{graphicx}
\usepackage{tabularx}
\usepackage{multirow}
\usepackage{indentfirst}
\usepackage[ruled,vlined]{algorithm2e}
\usepackage{xcolor}

\graphicspath{ {./figs/} }

\begin{document}

\begin{frontmatter}

\title{ Reinforcement Learning Provides a Flexible Approach for Realistic Supply Chain Safety Stock Optimisation}
\author[inst1]{Edward Elson Kosasih}
\author[inst1]{Alexandra Brintrup}
\affiliation[inst1]{organization={Institute for Manufacturing, University of Cambridge},country={United Kingdom}}
            



\begin{abstract}
Although safety stock optimisation  has been studied for more than 60 years, most companies still use simplistic means to calculate necessary safety stock levels, partly due to the mismatch between existing analytical methods' emphases on deriving provably optimal solutions and companies' preferences to sacrifice optimal results in favour of more realistic problem settings. A newly emerging method from the field of Artificial Intelligence (AI), namely Reinforcement Learning (RL), offers promise in finding optimal solutions while accommodating more realistic problem features. Unlike analytical-based models, RL treats the problem as a black-box simulation environment mitigating against the problem of oversimplifying reality. As such, assumptions on stock keeping policy can be relaxed and a higher number of problem variables can be accommodated. While RL has been popular in other domains, its applications in safety stock optimisation remain scarce. In this paper we investigate three RL methods, namely, Q-Learning, Temporal Difference Advantage Actor-Critic and Multi-agent Temporal Difference Advantage Actor-Critic for optimising safety stock in a linear chain of independent agents. We find that RL can simultaneously optimise both safety stock level and order quantity parameters of an inventory policy, unlike classical safety stock optimisation models where only safety stock level is optimised while order quantity is predetermined based on simple rules. This allows RL to model more complex supply chain procurement behaviour. However, RL takes longer time to arrive at solutions, necessitating future research on identifying and improving trade-offs between the use of AI and mathematical models are needed. 
\end{abstract}


\end{frontmatter}

\section{Introduction}
Increasing supply chain complexity and economic uncertainty affects both supply and demand. Supply chain executives need to improve their inventory management strategy to handle emerging uncertainties (\cite{humair_incorporating_2013}). A 2011 report from Chief Supply Chain Officer (CSCO) Insights found that 76\% of surveyed executives listed inventory management excellence as either top priority or highly important (\cite{csco_five_2011}). 

In this paper, we focus on the problem of end-to-end safety stock placement, which is a subset of the general inventory optimisation problem. The objective of end-to-end safety stock placement is to calculate how much and where to keep extra stock across the whole supply chain to mitigate stockout risks due to supply and demand uncertainty.
 
The safety stock placement problem, also known as safety stock optimisation, has been studied for more than 60 years, starting from the two seminal papers of \citet{simpson_jr_-process_1958} and \citet{clark_optimal_1960}. Significant emphases have been put on deriving optimisation strategies with provable optimality, often at the expense of simplifying reality. Unfortunately, these limiting assumptions prevent industry adoption (\cite{humair_incorporating_2013}). The CSCO 2011 report found that many companies, in practice, tend to still use rather simplistic means to calculate safety stock levels (\cite{csco_five_2011}). This is partly because companies are typically not as concerned about using the most optimally proven yet oversimplified policy but instead would like models that can reflect their reality (\cite{humair_incorporating_2013}), pointing to a discrepancy between academic studies and practice.

We hypothesise that a recently emerging technique from the field of AI Reinforcement Learning (RL) could help address this requirement. In RL goal-directed agents learn policies that optimise the performance rewards they would receive from the black-box environment. The agents make limited assumptions about this black-box behaviour. The learning algorithm of RL is designed to be as generic as possible such that if the black-box environment is altered, no algorithmic design changes are required. The agents thus could utilise the same learning algorithm to adapt their behaviour to the new environment. In the case of inventory optimisation, the environment could be defined as the companies' supply chain while the agents represent their safety stock control software.

To test our hypothesis we examine the use of RL by contextualising the safety stock problem as Q-Learning, Temporal Difference Advantage Actor-Critic and Multi-agent Temporal Difference Advantage Actor-Critic RL approach. A baseline analytical model and a test problem are developed for cross-comparison of results. Our findings show that while RL is slightly suboptimal compared to the baseline results that are derived analytically, it is able to optimise both safety stock level and order quantity parameters of an inventory policy - requirement that is not possible to fulfill with classical safety stock optimisation modelling where only safety stock level is optimised whilst order quantity rule being a priori predetermined based on a simple rule (such as base stock, (s, S) or (s, nQ) inventory policy). On the other hand, we find that RL suffers from high computational complexity, thus the benefits obtained from RL must be balanced against the time required to obtain solutions.

This paper is organised as follows. First, a literature review is performed to identify existing safety stock optimisation methods and identify research gaps. RL is proposed as a potential method to address these problems. Second, we design a problem to test the RL methods and solve it analytically. Next, we elaborate on the RL algorithmic design used to tackle the problem. Experimental results with the RL algorithms are analysed and discussed, followed by suggestions for future avenues of research.

\section{Literature Review}

The simplest formulation of the safety stock placement problem is the following: Minimise the total safety stock costs across all echelons in a supply chain, whilst ensuring that stockouts are prevented. This problem has been studied for more than 60 years, starting from the two seminal papers (\citet{simpson_jr_-process_1958} and \citet{clark_optimal_1960}). Over the years, safety stock placement has evolved to include more variables, capturing the various intricacies of supply chains. 

We perform a literature review on safety stock optimisation papers from 1958 until early 2020. Two electronic databases were searched, Scopus and Web of Science using the following keywords.

\begin{quote}
\centering
    "safety stock optimisation" \textbf{OR} "safety stock placement"
\end{quote}

Several comprehensive surveys on safety stock optimisation have been written by \cite{diks_multi-echelon_1996}, \cite{simchi-levi_performance_2011}, \cite{eruguz_comprehensive_2016}, and \cite{de_kok_typology_2018}. There are two classes of methods in the literature, Guaranteed Service Model (GSM) and Stochastic Service Model (SSM), introduced by \cite{simpson_jr_-process_1958} and \cite{clark_optimal_1960} respectively. GSM assumes that all demand will be fulfilled in fixed lead time, even if that means using extraordinary means like outsourcing or expediting if needed when inventory is insufficient. Meanwhile, SSM allows some demand to be back ordered when inventory is not enough, resulting in stochastic lead time from the customer's perspective.

\cite{graves_supply_2003} implemented the two models on several real-life problems and concluded that more empirical studies are needed to determine which models provide better and more realistic decision support. \cite{klosterhalfen_safety_2010} compared the two models and concluded that GSM performs slightly better than SSM given moderate cost for flexibility measures (such as outsourcing or expediting), large processing times in warehouse and high service level requirement for retailers. \cite{eruguz_comprehensive_2016} found that research on GSM has especially attracted more interest in the past decade. Following this trend, we focus on GSM in this work.

The most simplistic GSM formulates the inventory of a particular agent $j$ as a function of mean $\mu_j$ and standard deviation $\sigma_j$ of demand (modelled as a Normal distribution), supplier's lead time $SI_j$, processing time $T_j$, service time $S_j$, service level $z_j$ and Safety Stock function $SS_j()$ as seen in equation \ref{eq:inventory-ss}.

\begin{align}
\label{eq:inventory-ss}
I_j = \mu_j \times (SI_j + T_j - S_j) + SS_j(z_j, \sigma_j, SI_j, T_j, S_j)
\end{align}

Since mean $\mu$ demand is a fixed value, only the safety stock part of the equation is considered for optimisation. Given that $h_j$ represents inventory holding cost at agent $j$, the optimisation problem can be formulated as a minimisation problem over all $N$ agents in the supply chain, as seen in equation \ref{eq:ss-objective}.

\begin{align}
\shortintertext{Minimise}
\label{eq:ss-objective}
\sum_{j \in N} h_j \times SS_j(z_j, \sigma_j, SI_j, T_j, S_j) \\
\shortintertext{subject to}
\label{eq:ss-constraint-S-SI-T}
S_j - SI_j \leq T_j \text{ for all } j \in \text{nodes } \textbf{N}, \\
SI_j - S_i \geq 0 \text{ for all } (i, j) \in \text{set of adjacent nodes }  \textbf{A} \\
S_j \leq s_j \text{ for all } j \in \text{set of demand nodes } \textbf{D} \\
S_j, SI_j \geq 0 \text{ and integer for } j \in \text{nodes } \textbf{N} 
\end{align}

The definition of safety stock function $SS_j$ varies across the literature. The differences mainly come from modifying the following safety stock formula originally proposed by \cite{simpson_jr_-process_1958}.

\begin{align}
SS_j(z_j, \sigma_j, SI_j, T_j, S_j) = z_j \times \sigma_j \times \sqrt{SI_j + T_j - S_j}
\label{eq-general-SS}
\end{align}

where $z_j$ = service level, or multiplier for standard deviation $\sigma_j$ of demand seen at node $j$; this affects how much demand bound would be covered, $SI_j$ = service time of node $j$'s supplier, $T_j$ = processing time and $S_j$ = service time guaranteed to customer. Hence, safety stock is defined as inventory used to cover the lead time period ($SI_j + T_j - S_j$).

The following are several variations of the safety stock formula in equation \ref{eq-general-SS} proposed in the literature. \citet{inderfurth_safety_1991} models a case where there might be multiple end customers with correlated demands.  \citet{schoenmeyr_strategic_2009} assumes that future demand can be forecasted at present time by some function. \citet{graves_strategic_2016} adds a new variable called capacity constraints. \citet{graves_strategic_2008} works with nonstationary demand, where planning horizon can be divided into two phases each having different probability distribution. \citet{klosterhalfen_strategic_2014} consider a problem version where there is static dual supply. Finally, \citet{humair_optimizing_2006} defines a new attribute called clusters of commonalities.

To solve this problem formulation, most literature used analytical methods / mathematical programming approaches. \citet{simpson_jr_-process_1958} theoretically proved that in a serial supply chain, the problem is equivalent to a square root minimisation function over a bounded polyhedron. Therefore, the optimal solution lies at the vertices of the solution set. One can then enumerate all vertices and find the optimal result. This property is also referred to as an all-or-nothing characteristic. \citet{inderfurth_safety_1991} proved that the same property holds for the distribution network. Besides enumeration, \citet{graves_optimizing_2000}, \citet{inderfurth_safety_1991}, and \citet{minner_strategic_2012} have also shown that a dynamic programming algorithm can be used to find the solution iteratively. \citet{lesnaia_optimizing_2004} combined this with a branch-and-bound algorithm. \citet{magnanti_inventory_2006} estimates the objective to successive piece-wise linear function approximation, converting this into a sequence of linear programming problems. Heuristics have also been studied. \citet{shu_efficient_2009} uses linear approximations heuristics. \citet{minner_strategic_2012} investigates simulated annealing and tabu search. \citet{li_new_2012} uses hybrid constraint programming and genetic algorithms. 

Simulation-based optimisation methods have not been explored as much as analytical approaches in the literature. Our finding agrees with \cite{de_kok_typology_2018}, that simulation based approaches have been understudied in the wider literature of inventory optimisation. Unlike analytical models, simulation-based optimisation is done with respect to a black box. The problem's objectives and constraints are hidden within the latter, hence optimisation algorithms can only see  total costs of certain input configurations. In practice, this means that companies can use their ERP system as a black box, without the need to write down every subsystem interaction in a set of explicit equations. This is why \cite{klemmt_simulation-based_2009} wrote that simulation-based models require less modeling effort than their analytical model counterparts. According to \citet{glasserman_sensitivity_1995}, analytical methods focus on modeling more complex features while simulation-based methods accept the limitations of tractable models and adjust them accordingly to handle practical environments based on real system.

\begin{figure}
\centering    
\includegraphics[width=0.5\textwidth]{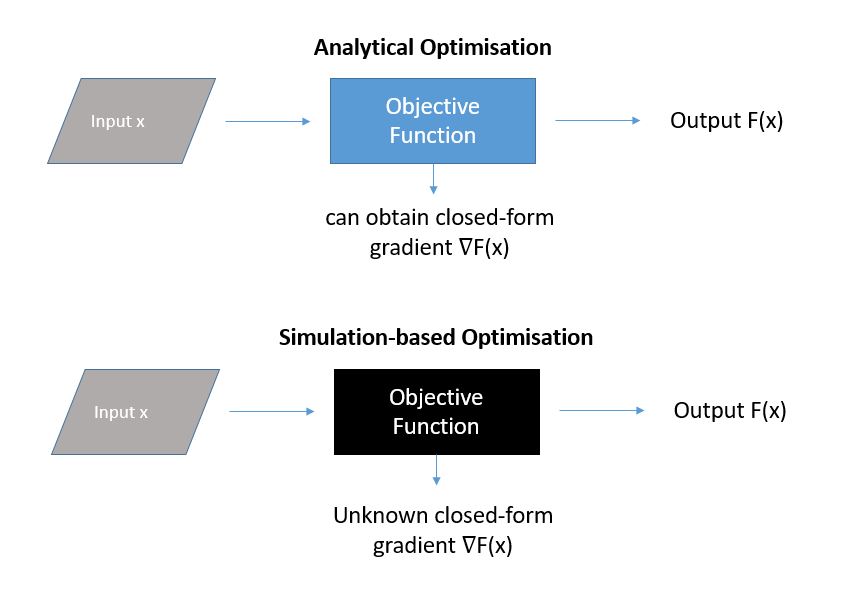}
\caption[Analytical vs Simulation-based Optimisation]{In analytical optimisation, the objective function is known in closed-form equation, hence we can obtain its gradient and solve it with mathematical programming. In simulation-based optimisation, the objective function is a black box, hence all we get is an output response for every input. We can not derive its closed-form gradient, thus we need to rely on approximations. While closed-form objective is generally preferable, in complex supply chain this might not always be obtainable, unless further assumptions and simplifications are made.}
\label{fig:analytical-vs-simulation-design}
\end{figure}

Some existing works on simulation-based optimisation in the literature included the following. \citet{sitompul_safety_2008} used Monte Carlo simulation to estimate safety stock correction factor $\phi_j$ that will then be fed into the analytical safety stock formulation. \citet{glasserman_sensitivity_1995} combined analytical- and simulation- based optimisation, where the latter is used to run sensitivity analysis and adjust the solution proposed by the former subsequently. \citet{schoenmeyr_strategic_2008} used numerical simulations to estimate upper limit of demand in the absence of explicit capacity constraints. Simulation shows it is optimal to set the upper limits far downstream in the supply chain using limits slightly larger than the average demand. \cite{molinder_joint_1997} used simulated annealing as a hybrid simulation-optimisation procedure to optimise safety stock and order quantity in a given MRP system. \citet{agarwal_multi-echelon_2019} used Gaussian process under a simulation-optimisation framework for inventory control.

Existing safety stock optimisation models also assume that the procurement rule of an inventory policy i.e. order quantity, was predetermined based on a simple rule. Only the safety stock level is optimised. Three of the most commonly used inventory policies are base stock, (s, S) and (s, nQ) (\cite{de_kok_typology_2018}). In base stock, agents continuously review their inventory position and upon demand fulfillment, place an order that will bring the latter up to predetermined base stock level. (s, S) is almost identical to the former, where base stock level is denoted as S. However, an order will only be placed if the inventory position has decreased below reorder point s. Lastly, in (s, nQ) orders can only be an integer $n$ multiplication of fixed quantity Q, and are placed if a reorder point $s$ has been achieved. These policies can be proven to be optimal for standard problems. However, for a black box complex problem that changes over time, the order quantity rule aspect of inventory policy might have to be optimised as well, and not determined beforehand. 

\begin{figure} 
\centering    
\includegraphics[width=0.5\textwidth]{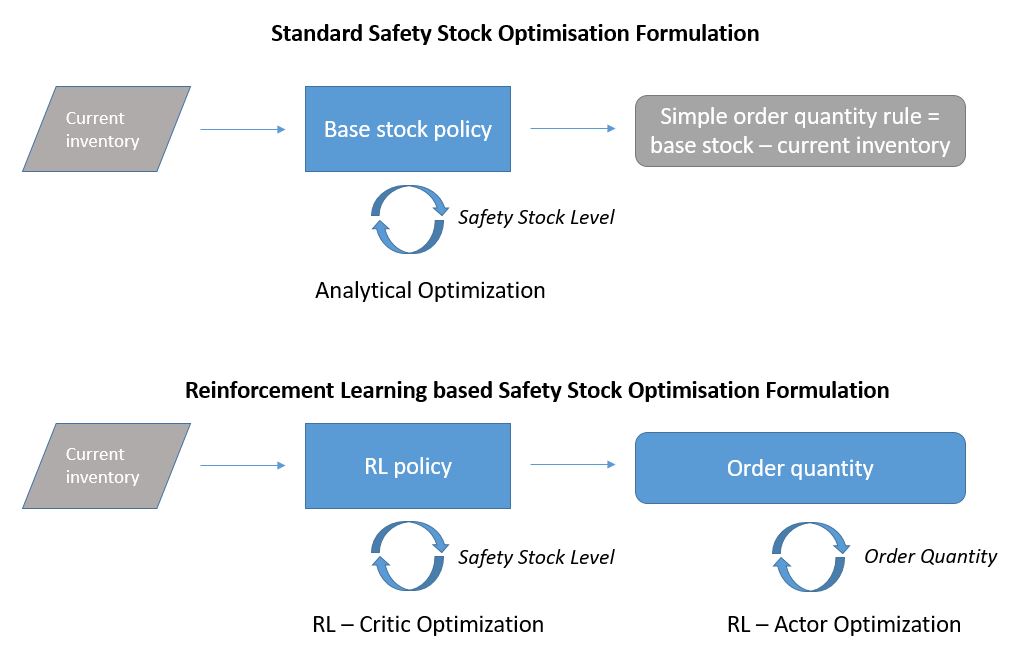}
\caption[RL-based Optimisation]{In standard safety stock optimisation models, only the safety stock level is optimised. Meanwhile, the order quantity was predetermined based on a simple rule. For example, in base stock policy, the order quantity rule is set beforehand to be equal to base stock - current inventory. On the other hand, in RL-based models, both safety stock level and order quantity rule are optimised. In Actor-Critic model, for instance, safety stock level is modeled with a critic neural network while order quantity with actor neural network.}
\label{fig:current-vs-rl-policy}
\end{figure}

We hypothesise that a recently developed family of algorithms, called Reinforcement Learning (RL), could address the two aforementioned research gaps i.e. lack of simulation-based optimisation methods and predetermined inventory policy. RL is an area in machine learning where agents learn how to map situations to actions in order to maximise/minimise a certain reward signal provided by the environment (\cite{sutton_reinforcement_2018}). RL agents are inherently trained to learn from a black box environment (equivalent to simulation-based models), discover both the optimal final state and also the policy (or set of actions) that must be taken to achieve that state (i.e. optimising inventory policy). Figure \ref{fig:current-vs-rl-policy} shows the difference between current safety stock optimisation model and our proposed RL approach.

RL has recently been shown to be successful in tackling certain complex problems that were once thought intractable, such as the game of Go  (\cite{vinyals_grandmaster_2019}, \cite{silver_mastering_2016}, \cite{silver_mastering_2017}, \cite{silver_reinforcement_2009}). However,  there is no existing research that used Reinforcement Learning as a methodology for safety stock optimisation. As such, we expanded our literature review to the wider inventory control problem. After doing so, we found a limited number of papers as reviewed below.

\citet{oroojlooyjadid_deep_2017} uses RL to solve the beer distribution game. They use Deep Q-Network (DQN) algorithm to train an agent that could adjust its ordering policy in a simulated Beer Distribution Game i.e. a multi-echelon supply chain problem. They found that DQN could replicate base-stock policy and also learn more complex policies when faced with an atypical environment. \citet{kara_reinforcement_2018} uses RL to specify ordering policies of perishable inventory systems. The authors found that using Q-Learning and SARSA produces better results when demand has high variance and products have short lifetimes. \citet{gijsbrechts_can_2019} uses Asynchronous Advantage Actor-Critic (A3C) algorithm to solve Dual Sourcing Inventory Model, Lost Sales Inventory Models and Multi-echelon Inventory Model problem. They concluded that RL is a promising generic technique that can be used to solve intractable inventory problems, although it does not always outperform other algorithms. Moreover, interpreting the resulting trained policy remains a challenge. \citet{jiang_case-based_2009} combines case-based RL with multi-agent system to control inventory level while making sure that the target service level is met. The agents managed to learn the parameters of (s, S) and (s, nQ) inventory policies. 

Nevertheless, none of the aforementioned works addressed and benchmarked against existing analytical-based safety stock optimisation models. In this work, we show how RL can be used to solve safety stock optimisation problem. While RL is slightly suboptimal compared to the analytical-based model, it could produce a more complex inventory policy as both safety stock level and order quantity rule are optimised. To illustrate how RL-based model works in practice, we explore three commonly used RL algorithms: Q-Learning (Algorithm \ref{algo:q-learning}), Temporal Difference Advantage Actor-Critic (Algorithm \ref{algo:centralised-a2c}) and Multi-agent Temporal Difference Advantage Actor-Critic (Algorithm \ref{algo:decentralised-a2c}) (see Appendix). The next section introduces our problem design. 

\section{Problem Design}

The objective of our experiment is to juxtapose the insights that existing analytical-based model and our proposed RL-based model yield. Therefore, in order to make comparable results, we resorted to a standard serial supply chain problem, whose analytical solutions have been thoroughly studied and the optimal solution can be derived and proven. We shall show how reinforcement learning can optimise both safety stock level and order quantity rule through the usage of critic and actor neural network, respectively.

\begin{figure}[htbp!] 
\centering    
\includegraphics[width=0.5\textwidth]{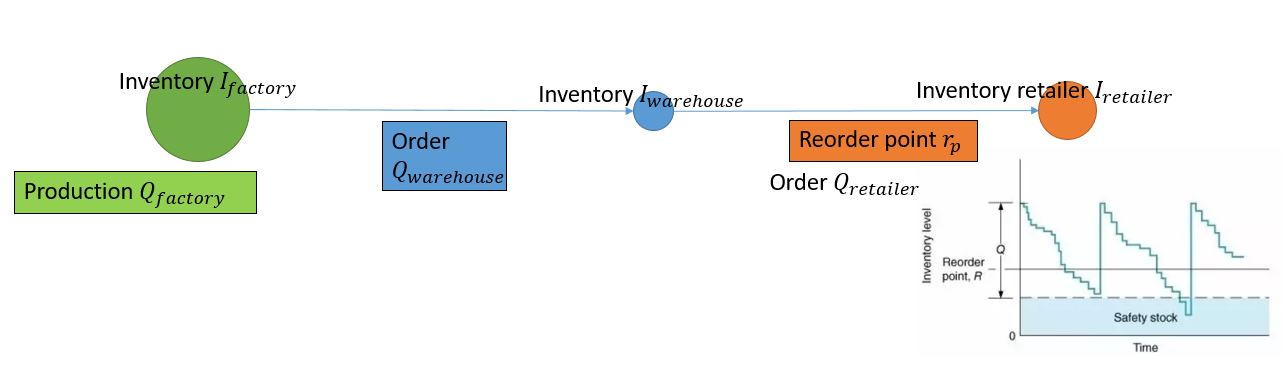}
\caption[Simulation Design]{Simulation Design}
\label{fig:sim-design}
\end{figure}

There are 3 agents in the three-echelon supply chain: a Retailer, warehouse and factory. Retailer adopts a (Q, $r_p$) inventory policy with $Q_{retailer}$ being normally distributed order quantity that is placed to warehouse when its’ inventory position hits reorder point $r_p$. Being a pull-based supply chain, Factory and Warehouse will only procure or produce stocks upon seeing demand from the Retailer. Factory and Warehouse could choose to order or produce more than demand, so that any extra items (also known as safety stock) are kept as inventories in the next cycle. With more inventories on-hand, Factory and Warehouse could offer a shorter service time to their customers. Any demand that cannot be met (e.g. when actual fulfillment exceeds guaranteed service time) will be deemed as stock-out.

The objective of the problem is to minimise the amount of safety stock in the system while ensuring that there is no stock-out. We assume a cooperative game setup where agents work together to optimise a common objective function. Moreover, we would like to extract insights on how agents’ ordering policy should be in order to navigate through the transient state and achieve the optimal inventory position as prescribed by the GSM. 

As we will use reinforcement learning for this experiment, we need to define three set of variables: agent states, agent actions, and environment model. Two sets of experiments will be done on different cost assignments. The problem definition can be found in Table \ref{ref:problems-def}. This table is divided into 5 sections, combining optimisation problem (variables, objective, constraints) and Markov Decision Problem (state, action, environment) formulation.

Using the inventory optimisation model typology developed by \cite{de_kok_typology_2018}, our proposed method belongs to category 3, S, D, G $\mid$ F, D $\mid$ N, G, $\mid$ O,  F, U $\mid$ M $\mid \mid$ S, O.

\begin{center}
\begin{table*}

\centering
\caption*{(a) Problem State Variables}
\begin{tabular}{ | m{4cm}| m{1.5cm} | } 
\hline
Inventory at Factory & $I_{factory}$ \\ 
\hline
Inventory at Warehouse & $I_{warehouse}$ \\ 
\hline
\end{tabular}

\bigskip
\centering
\caption*{(b) Problem Action Variables}
\begin{tabular}{ | m{7cm}| m{3cm} | } 
\hline
Order Quantity from Retailer to Warehouse & $Q_{retailer}$ \\
\hline
Reordering Point of Retailer & $r_p$ \\
\hline
Order Quantity from Warehouse to Factory & $Q_{warehouse}$ \\
\hline
Production Quantity of Factory & $Q_{factory}$ \\
\hline
\end{tabular}

\bigskip

\centering
\caption*{(c) Problem Environment Variables}
\begin{tabular}{ | m{5cm}| m{3cm} | m{4cm} | } 
\hline
Inventory Cost of Factory & $h_{factory}$ & £ $h_{factory}$/item \\ 
\hline
Inventory Cost of Warehouse & $h_{warehouse}$ & £ $h_{warehouse}$/item \\ 
\hline
Processing Times of Factory & $T_{factory}$ & 1 time units \\
\hline
Processing Times of Warehouse & $T_{warehouse}$ & 3 time units \\
\hline
Service Time of Retailer & $S_{retailer}$ & 0 time units \\
\hline
Demand Profile & D & $Normal(2, 0.01)$ \\
\hline
\multirow{3}{*}{Maximum Inventory Capacity}
&$C_{factory}, C_{warehouse}$ & 30 items\\
&$C_{retailer}$ & \\ 
\hline
Stock-out Cost at Retailer & $\eta$ & £10000/item \\
\hline
Order Quantity from Retailer to Warehouse & $Q_{retailer}$ & $Normal(\mu = 10, \sigma = 1)$ items \\
\hline
Target Service Level z-score & $z_{factory}, z_{warehouse}$ & 3, or 99.9\% service level  \\
\hline
\end{tabular}

\bigskip

\centering
\caption*{(d) Problem Constraints}
\begin{tabular}{ | m{6cm} | m{6cm} | } 
\hline
$r_p$ $\geq$ 0 & reorder point is non-negative \\
\hline
$r_p$ $\leq$ 6 & trivial solution is not possible. With demand $\mu=2$, this means that $S_{warehouse} \leq \frac{r_p}{\mu} = \frac{6}{2} = 3 $ \\
\hline
$Q_{warehouse}$ $\geq$  $Q_{retailer}$ – $I_{warehouse}$ & new inventory after receiving order is larger than demand; hence customer’s order is fulfill-able. \\
\hline
$Q_{warehouse}$ $\leq$ $C_{warehouse}$ – $I_{warehouse}$ & new inventory after receiving order is less than capacity. \\
\hline
$Q_{factory}$ $\geq$ $Q_{warehouse}$ – $I_{factory}$ & new inventory after receiving order is larger than demand; hence customer’s order is fulfill-able. \\
\hline
$Q_{factory}$ $\leq$ $C_{factory}$ – $I_{factory}$ & new inventory after receiving order is less than capacity. \\
\hline
$r_p$, $Q_{warehouse}$ and $Q_{factory}$ & all integers \\
\hline
\end{tabular}

\bigskip

\centering
\caption*{(e) Problem Setup}
\begin{tabular}{ | m{2cm} | m{6cm} | } 
\hline
Case 1 & $h_{factory}$ = £1000 $\geq$ $h_{warehouse}$ = £5 \\ 
\hline
Case 2 & $h_{factory}$ = £5 $\leq$ $h_{warehouse}$ = £1000 \\ 
\hline
\end{tabular}

\bigskip
\caption{Problem Definition}
\label{ref:problems-def}
\end{table*}
\end{center}

\section{Analytical-based Optimisation}
In this section, we formulate this problem as a concave minimisation over bounded polyhedron. According to the formulation, the GSM problem is:

\begin{align}
\shortintertext{Minimise}
\begin{aligned}
h_{factory} \times z_{factory} \times \sigma \times \sqrt{0 + 1 - S_{factory}} + \\ h_{warehouse} \times z_{warehouse} \times \sigma \times \sqrt{S_{factory} + 3 - S_{warehouse}} \\
\end{aligned}
\shortintertext{subject to}
S_{factory} \leq 0 + 1 \\
S_{warehouse} \leq S_{factory} + 3 \\
S_{warehouse} \leq 3
\end{align}

where $z_j$ = service level = 3 and $\sigma$ = standard deviation of demand = 1 in our case. Therefore, the amount of safety stock kept in Factory is $ 3 \times 1 \times \sqrt{1-S_{factory}}$ while Warehouse keeps $3 \times 1 \times \sqrt{S_{factory}+3-S_{warehouse}}$. Based on equation \ref{eq:inventory-ss}, Total Inventory for both agents can be written as a sum of mean demand and safety stock. $I_{factory} = \mu \times (1-S_{factory}) + 3 \times 1 \times \sqrt{1-S_{factory}} $ and $I_{warehouse} = \mu \times (S_{factory}+3-S_{warehouse}) + 3 \times 1 \times \sqrt{S_{factory}+3-S_{warehouse}} $. 

The original GSM problem assumes base-stock (order up-to-Q) policy with continuous review. Following the all-or-nothing characteristic of GSM for serial supply chains, we can test the four vertices of the solution set, as seen in Table \ref{table:case-enumeration}. The optimal solution with the lowest cost is highlighted in blue.

\subsubsection{Case 1 Hypothesis: $S_{factory} = 1, S_{warehouse} = 3$, or $I_{factory} = 0, I_{warehouse} = 13, r_p = 6$}
Table \ref{table:case-enumeration}(a) shows that it is always better to keep no stock in Factory ($S_{factory}=1$, $I_{factory} = 0$). As we have set the problem constraint such that $S_{warehouse} <= 3$, the optimum solution should thus be when $S_{warehouse} = 3$, $I_{warehouse} = 13$. With $S_{warehouse} = 3$, agent Retailer's reorder point $r_p$ is then equal to the daily demand mean it sees multiplied by service time $r_p = S_{warehouse} \times D = 3 \times 2 = 6$.

\subsubsection{Case 2 Hypothesis: $S_{factory} = 0, S_{warehouse} = 3$, or $I_{factory} = 13, I_{warehouse} = 0, r_p = 6$}
Table \ref{table:case-enumeration}(b) shows that it is always better to keep no stock in Warehouse ($S_{warehouse}=3$, $I_{warehouse} = 0$). The optimum solution should thus be when $S_{factory} = 0$, $I_{factory} = 13$. With $S_{warehouse} = 3$, agent Retailer's reorder point $r_p$ is then equal to the daily demand mean it sees multiplied by service time $r_p = S_{warehouse} \times D = 3 \times 2 = 6$.

\begin{center}
\begin{table}[ht]
\centering
\caption*{(a) Solving Case 1 with Enumeration}
\begin{tabular}{ | m{2cm} | m{3cm} | m{2cm} | } 
\hline
Service Level & Inventory Level & Total Safety Stock Cost \\
\hline
$S_{factory} = 0$, $S_{warehouse} = 0$  & $I_{factory} = 13 $, $I_{warehouse} = 30 + 3\sqrt{3}$  & $3000 + 15 \sqrt{3}$ \\ 
\hline
$S_{factory} = 0$, $S_{warehouse} = 3$  & $I_{factory} = 13 $, $I_{warehouse} = 0$  & $3000$ \\ 
\hline
$S_{factory} = 1$, $S_{warehouse} = 0$  & $I_{factory} = 0 $, $I_{warehouse} = 46$  & $30$ \\ 
\hline
\color{blue} $S_{factory} = 1$, $S_{warehouse} = 3$  & \color{blue} $I_{factory} = 0 $, $I_{warehouse} =13$  & \color{blue} $15$ \\ 
\hline
\end{tabular}

\bigskip

\centering
\caption*{(b) Solving Case 2 with Enumeration}
\begin{tabular}{ | m{2cm} | m{3cm} | m{2cm} | } 
\hline
Service Level & Inventory Level & Total Safety Stock Cost \\
\hline
$S_{factory} = 0$, $S_{warehouse} = 0$  & $I_{factory} = 13 $, $I_{warehouse} = 30 + 3\sqrt{3}$  & $15 + 3000\sqrt{3}$ \\ 
\hline
\color{blue} $S_{factory} = 0$, $S_{warehouse} = 3$  & \color{blue} $I_{factory} = 13 $, $I_{warehouse} = 0$  & \color{blue} $15$ \\ 
\hline
$S_{factory} = 1$, $S_{warehouse} = 0$  & $I_{factory} = 0 $, $I_{warehouse} = 46$  & $6000$ \\ 
\hline
$S_{factory} = 1$, $S_{warehouse} = 3$  & $I_{factory} = 0 $, $I_{warehouse} = 13$  &  $3000$ \\
\hline
\end{tabular}

\bigskip

\caption{Solving Cases with Enumeration}
\label{table:case-enumeration}
\end{table}
\end{center}

Observe that the analytical-based method provides the optimal safety stock allocation i.e. how much inventory should be kept in each location. However, it does not simultaneously optimise the order quantity rule that the agents must follow in order to achieve that inventory position. Instead, it is assumed beforehand \textit{a priori} that base stock policy is used e.g. order quantity = base stock - current inventory. For a simple case such as this example problem, it has indeed been mathematically proven that base stock is the most optimum policy (\cite{eruguz_comprehensive_2016}). However, this is not true in general. Thus, this analytical formulation would not work for a more complex problem as we can no longer assume that base stock policy is optimal. In the next section, we shall see that RL allows simultaneous optimisation of order quantity rule and safety stock level, hence providing a way around this limitation.

\section{Simulation-based Optimisation with RL}

The way our simulated problem works is the following. The simulation consists of cycles of agents shipping or producing items upon seeing demands from their customers. For every cycle, after demand has been fulfilled, the following joint reward is provided to all agents. The higher the amount of inventory and stock-outs, the more negative will the reward be. The objective is thus to maximise the reward.

\begin{align}
\begin{aligned}
    \text{reward = } - \{ h_{factory} \times I_{factory} + h_{warehouse} \times I_{warehouse} \\ + \eta \times \text{stockouts} \}
\end{aligned}
\end{align}

\subsubsection{Tabular Q-Learning}

The planner agent sees a state vector $s_t = [ I_{factory} I_{warehouse} r_p^t ]$ and decides an action vector $a_t = \begin{bmatrix} Q_{factory} & Q_{warehouse} & r_p^{t+1} \end{bmatrix}$ accordingly. The planner agent is trained with \textbf{Algorithm \ref{algo:q-learning}} (see Appendix) where  Q-function is represented with a lookup table with hyperparameters specified in Table \ref{table:centralised-q-hyperparameters}.

\begin{center}
\begin{table}[ht]
\centering
\caption{Q-Learning Hyperparameters}
\begin{tabular}{ | m{2.5cm} | m{4.8cm} | } 
\hline
 & \textbf{Q-Table} \\
\hline
input vector & $s_t = \begin{bmatrix} I_{factory} & I_{warehouse} & r_p^t \end{bmatrix}$ \\
\hline
output vector & $a_t = \begin{bmatrix} Q_{factory} & Q_{warehouse} & r_p^{t+1} \end{bmatrix}$ \\
\hline
number of instances & 1 \\
\hline
discount factor $\gamma$ & 0.2 \\ 
\hline
learning rate $\alpha$ & 0.8 \\ 
\hline
epsilon-greedy factor $\epsilon$ & 0.5 \\ 
\hline
implementation & numpy \\ 
\hline
\end{tabular}
\label{table:centralised-q-hyperparameters}
\end{table}
\end{center}

\subsubsection{Temporal Difference Advantage Actor-Critic}

Although Q-Learning works, it does not scale well with the size of the action space as it assumes that actions are discrete. Moreover, it does not generalise across states and actions since it only uses tabular representation. In this section, we explore another approach called policy gradient that could operate on continuous space and generalise.

Training is done using \textbf{Algorithm \ref{algo:centralised-a2c}} (see Appendix). Both critic and actor agents are represented with neural networks whose hyperparameters are specified in Table \ref{table:centralised-a2c-hyperparameters}. We chose default hyperparameters that are prescribed by the tensorflow-keras software package. During training, both critic and actor are trained iteratively. However during testing, only the actor is used.

\begin{center}
\begin{table}
\centering
\caption{A2C Neural Network Hyperparameters}
\begin{tabular}{ | m{1.4cm} | m{2.9cm} | m{3.3cm} | } 
\hline
 & \textbf{Value Network} & \textbf{Policy Network} \\
\hline
number of instances & 1 & 1 \\
\hline
input & \small $s_t = \begin{bmatrix} I_{factory} & I_{warehouse} & r_p^t \end{bmatrix}$ & \small $s_t = \begin{bmatrix} I_{factory} & I_{warehouse} & r_p^t \end{bmatrix}$ \\
\hline
hidden layers and neurons & 3 layers each 100 neurons with ReLU activation & 3 layers each 100 neurons with ReLU activation \\
\hline
output & scalar score & \small $a_t = \begin{bmatrix} Q_{factory} & Q_{warehouse} & r_p^{t+1} \end{bmatrix}$ \\
\hline
discount factor $\gamma$ & 0.2 & 0.2 \\ 
\hline
optimiser & Adam & Adam \\ 
\hline
optimiser's $\alpha$ & 0.001 & 0.001 \\ 
\hline
optimiser's $\beta_1$ & 0.9 & 0.9 \\ 
\hline
optimiser's $\beta_2$ & 0.999 & 0.999 \\ 
\hline
optimiser's $\epsilon$ & 1e-07 & 1e-07 \\ 
\hline
software & tensorflow-keras & tensorflow-keras \\ 
\hline
\end{tabular}
\label{table:centralised-a2c-hyperparameters}
\end{table}
\end{center}

\subsubsection{Multi-agent Temporal Difference Advantage Actor-Critic}

Although centralised methods work, they scale exponentially with the number of agents. In this section we explore decentralised methods where each agent can only see its' own inventory and demand. As a result, the model would scale linearly with the number of agents. In particular we focus on a particular strand of decentralised reinforcement learning algorithm called multi-agent A2C.

As mentioned in \textbf{Algorithm \ref{algo:decentralised-a2c}} (see Appendix), this method requires a centralised critic that will only be present during training. Meanwhile, in testing, the actors are all decentralised. 

The critic sees joint state \small $s_t = \begin{bmatrix} I_{factory} & I_{warehouse} &  r_p^t \end{bmatrix}$, while the actor sees a local state i.e. \small $s_t^{factory} = \begin{bmatrix} I_{factory} & Q_{warehouse} \end{bmatrix}$ for factory, and \small $s_t^{warehouse} = \begin{bmatrix} I_{warehouse} & Q_{retailer} \end{bmatrix}$ for warehouse respectively. Training is done using \textbf{Algorithm \ref{algo:decentralised-a2c}}. Both critic and actor are represented with neural networks whose hyperparameters are specified in Table \ref{table:decentralised-a2c-hyperparameters}. We chose default hyperparameters that are prescribed by the tensorflow-keras software package.

\begin{center}
\begin{table}
\centering
\caption{Multi-agent A2C Neural Network Hyperparameters}
\begin{tabular}{ | m{1.4cm} | m{2.9cm} | m{3.3cm} | } 
\hline
 & \textbf{Value Network} & \textbf{Policy Network} \\
\hline
number of instances & 1 & 3 \\
\hline
input & \small $s_t = \begin{bmatrix} I_{factory} & I_{warehouse} & r_p^t \end{bmatrix}$ & \small $s_t^i = \begin{bmatrix} I_i & Q_{i+1} \end{bmatrix}$ for $i \in \{factory, warehouse\}$ \\
\hline
hidden layers and neurons & 3 layers each 100 neurons with ReLU activation & 3 layers each 100 neurons with ReLU activation \\
\hline
output & scalar score & \small $a_t^i = Q_i$ for $i \in \{factory, warehouse\}$ and \small $a_t^{retailer} = r_p^{t+1}$ \\
\hline
discount factor $\gamma$ & 0.2 & 0.2 \\ 
\hline
optimiser & Adam & Adam \\ 
\hline
optimiser's $\alpha$ & 0.001 & 0.001 \\ 
\hline
optimiser's $\beta_1$ & 0.9 & 0.9 \\ 
\hline
optimiser's $\beta_2$ & 0.999 & 0.999 \\ 
\hline
optimiser's $\epsilon$ & 1e-07 & 1e-07 \\ 
\hline
software & tensorflow-keras & tensorflow-keras \\ 
\hline
\end{tabular}
\label{table:decentralised-a2c-hyperparameters}
\end{table}
\end{center}


\section{Experimental Results}

Experiments were performed on a Dell laptop with Intel i9-9980HK CPU, 2.4 GHz processor and 16 GB RAM.

\subsubsection{Tabular Q-Learning}

Table \ref{table:centralised-q-1-2-ci} shows the resulting 95\% CI average inventory level for both case 1 and case 2 when the agents are trained from scratch 10 times with different random initial conditions. As hypothesised, the result shows that the agent's behaviour is suboptimal but close to the analytically-derived solutions, as defined in the two columns: Case 1 and Case 2 Analytical.

\begin{center}
\begin{table}[ht]
\centering
\caption{Q-Learning Case 1 and 2 95\% CI Average Inventory Level}
\begin{tabular}{ | m{1.5cm} | m{1cm} | m{1cm} | m{1cm} | m{1cm} | } 
\hline
 & Case 1 RL & Case 1 Analytical & Case 2 RL & Case 2 Analytical \\
\hline
$r_p$ & [1.02, 1.12] & 6 & [3.58, 3.86] & 6 \\
\hline
$I_{warehouse}$ & [11.22, 13.34] & 13 & [1.29, 1.7] & 0 \\ 
\hline
$I_{factory}$ & [0.41, 0.98] & 0 & [11.91, 13.18] & 13 \\ 
\hline
\end{tabular}
\label{table:centralised-q-1-2-ci}
\end{table}
\end{center}

The results demonstrate that Q-Learning works well, however the main caveat is that this method assumes a discrete action space. In our problem case where each agent could choose between ordering 0 to 30 items (maximum capacity), this means that each agent has 30 possible actions. This scales exponentially with the number of agents (assuming n agents all with similar action space, the joint action space will be $30^n$). Moreover the agent does not generalise across states and actions as each entry is treated as another row in the table.

\subsubsection{Temporal Difference Advantage Actor-Critic}
From Table \ref{table:centralised-a2c-1-2-ci} we can observe the resulting 95\% CI average inventory level for both case 1 and case 2 when the agents are trained 10 times from different random initial conditions. The agents learned to keep inventory levels that as hypothesised, are suboptimal but close to the solutions that were derived analytically.

\begin{center}
\begin{table}[ht]
\centering
\caption{A2C Case 1 and 2 95\% CI Average Inventory Level}
\begin{tabular}{ | m{1.5cm} | m{1cm} | m{1cm} | m{1cm} | m{1cm} | } 
\hline
 & Case 1 RL & Case 1 Analytical & Case 2 RL & Case 2 Analytical \\
\hline
$r_p$ & [2.11, 2.84] & 6 & [3.18, 3.77] & 6 \\
\hline
$I_{warehouse}$ & [12.58, 15.11] & 13 & [0.37, 0.85] & 0 \\ 
\hline
$I_{factory}$ & [1.66, 2.87] & 0 & [8.97, 11.24] & 13 \\ 
\hline
\end{tabular}
\label{table:centralised-a2c-1-2-ci}
\end{table}
\end{center}

\begin{figure}[htbp!] 
\centering    
\includegraphics[width=0.5\textwidth]{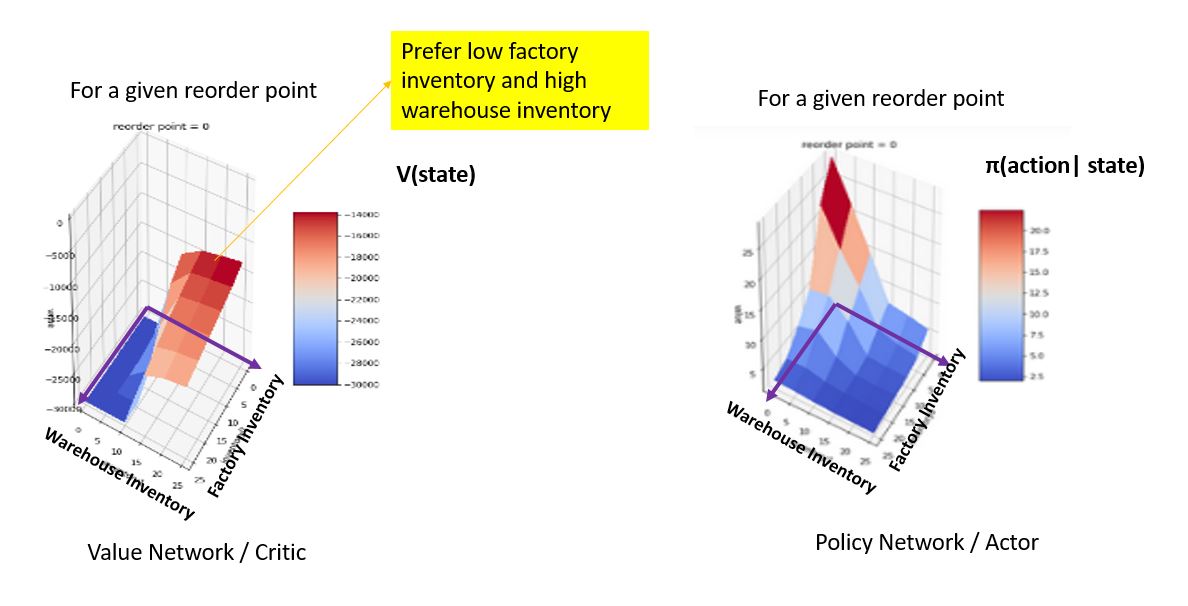}
\caption*{(a) A2C Case 1 Value-Policy Visualisation }

\bigskip

\centering    
\includegraphics[width=0.5\textwidth]{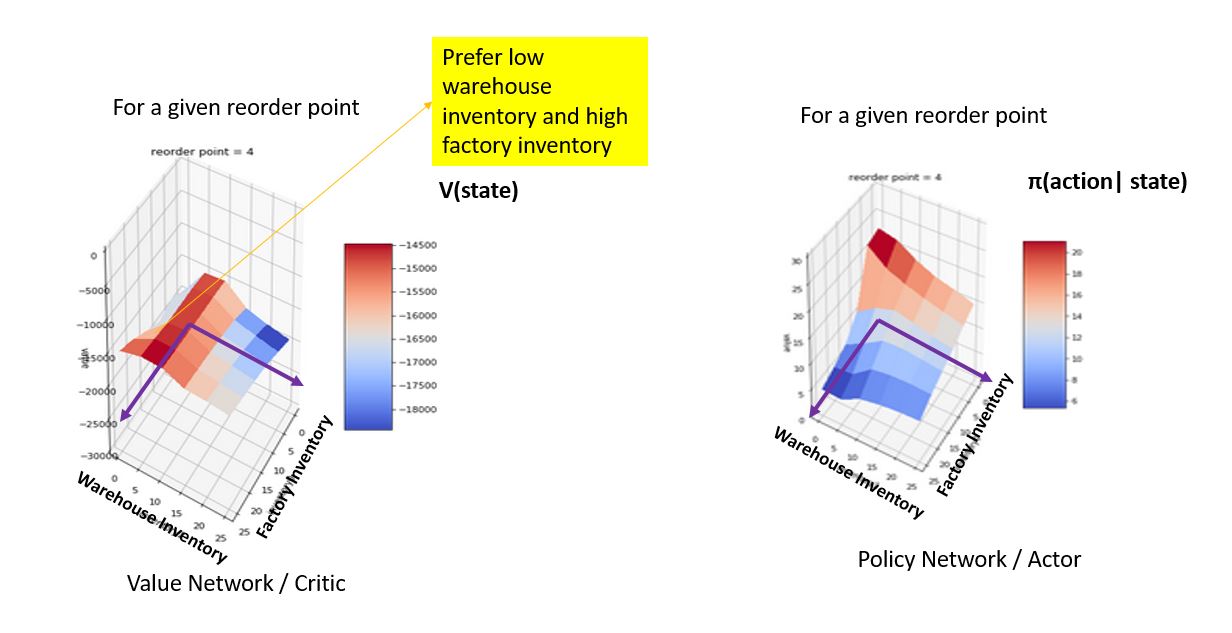}
\caption*{(b) A2C Case 2 Value-Policy Visualisation }

\bigskip

\caption{A2C Value-Policy Visualisation}
\label{fig:centralised-a2c-viz}
\end{figure}

Besides inventory level, A2C framework also provides additional insights as we can visualise both the actor's and the critic's behavior i.e. input-output relationship. We analyse both critic/value and actor/policy network's input-output mapping for both cases 1 (Figure \ref{fig:centralised-a2c-viz}(a)) and 2 (Figure \ref{fig:centralised-a2c-viz}(b)). 

Figure \ref{fig:centralised-a2c-viz}(a) indicates that state $I_{factory} = 0$ and $I_{warehouse} > 13$ are preferable for case 1, as predicted by our hypothesis. We've also plotted the policy network output for agent Factory. $Q_{factory}$ is always kept to a minimum unless $I_{factory}$ and $I_{warehouse}$ approach zero. This encourages Factory to fulfill demand from existing stocks, hence resulting in a lean behaviour. Meanwhile, Figure \ref{fig:centralised-a2c-viz}(b) shows that for case 2 state $I_{factory} > 13$ and $I_{warehouse} = 0$ are preferable, similar to our hypothesis. Also plotted is the policy network output for agent Factory. $Q_{factory}$ is always placed if $I_{factory}$ approaches zero. However, if $I_{factory}$ is too much, $Q_{factory}$ will be cut down to zero, encouraging Factory to fulfill demand from existing stocks. This results in a consistent, innovative stock-keeping behavior.

\subsubsection{Multi-agent Temporal Difference Advantage Actor-Critic}
Similar to Q-Learning and A2C, we train the agents 10 times from different random initial conditions for both cases 1 and 2, resulting in 95\% CI average inventory level as seen in Table \ref{table:decentralised-a2c-1-2-ci}.

\begin{center}
\begin{table}[ht]
\centering
\caption{Multi-agent A2C Case 1 and 2 95\% CI Average Inventory Level}
\begin{tabular}{ | m{1.5cm} | m{1cm} | m{1cm} | m{1cm} | m{1cm} | } 
\hline
 & Case 1 RL & Case 1 Analytical & Case 2 RL & Case 2 Analytical \\
\hline
$r_p$ & [2.73, 3.09] & 6 & [2.94, 3.46] & 6 \\
\hline
$I_{warehouse}$ & [13.98, 19.46] & 13 & [1.97, 2.96] & 0 \\ 
\hline
$I_{factory}$ & [0.58, 1.64] & 0 & [8.64, 10.98] & 13 \\ 
\hline
\end{tabular}
\label{table:decentralised-a2c-1-2-ci}
\end{table}
\end{center}

\begin{figure}[htbp!] 
\centering    
\includegraphics[width=0.5\textwidth]{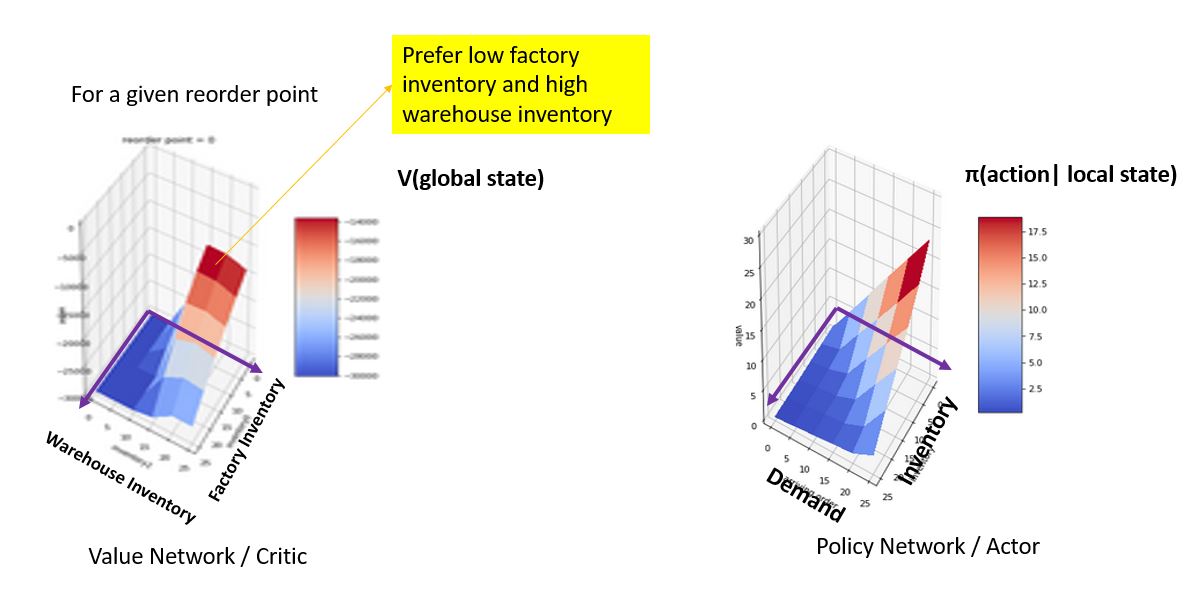}
\caption*{(a) Multi-agent A2C Case 1 Value-Policy Visualisation }

\bigskip

\centering    
\includegraphics[width=0.5\textwidth]{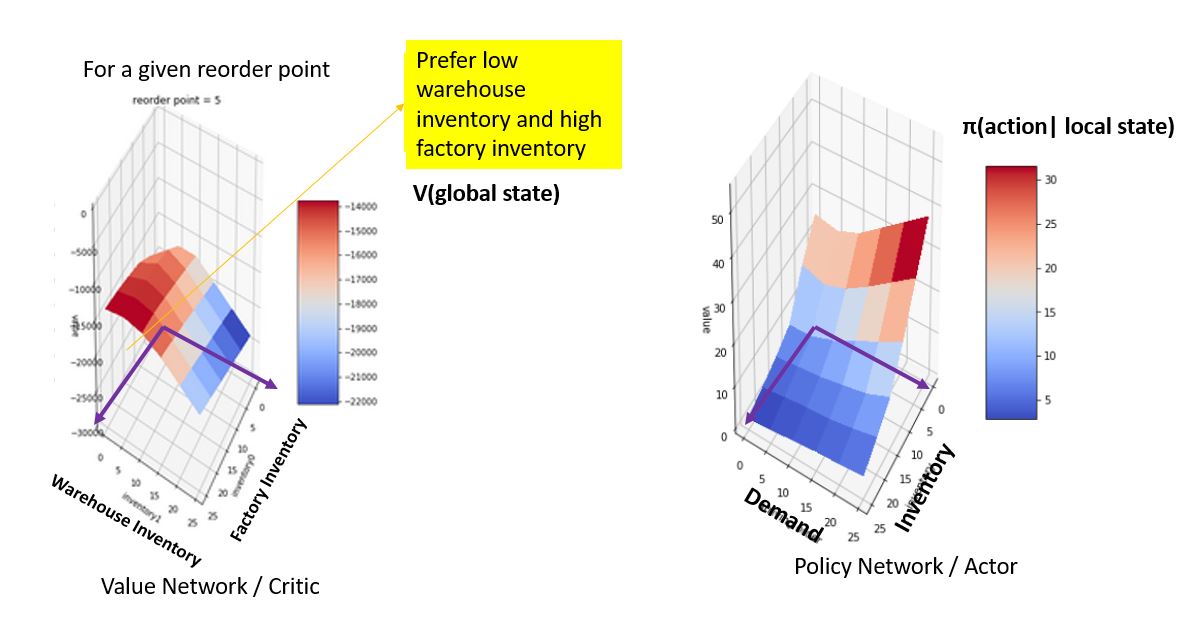}
\caption*{(b) Multi-agent A2C Case 2 Value-Policy Visualisation }

\bigskip

\caption{Multi-agent A2C Value-Policy Visualisation }
\label{fig:decentralised-a2c-viz}

\end{figure}

Multi-agent A2C, akin to the previous A2C algorithm, also provides additional insights as we can visualise both actor's and critic's behavior i.e. input-output relationship. These can be seen in Figure \ref{fig:decentralised-a2c-viz}(a) Figure \ref{fig:decentralised-a2c-viz}(b), for case 1 and 2 respectively. 

From Figure \ref{fig:decentralised-a2c-viz}(a), we can see that the agents prefer state $I_{factory} = 0$ and $I_{warehouse} > 13$, as predicted by our hypothesis. Looking at the policy network output for agent Factory, we can obtain a better understanding of the agent's behaviour. In this case, Factory will only produce $Q_{factory}$ if $I_{factory} = 0$ and it receives demand $Q_{warehouse}$ from Warehouse. This results in lean behaviour. On the other hand for case 2, Figure \ref{fig:decentralised-a2c-viz}(b) shows that state $I_{factory} > 13$ and $I_{warehouse} = 0$ are preferable. We've also plotted the policy network output for agent Factory. Factory will always produce $Q_{factory}$ if $I_{factory}$ approaches zero, regardless of demand. However, if $I_{factory}$ is too much, $Q_{factory}$ will be cut down to a minimum, encouraging Factory to use existing stocks to fulfill demand. Once more this results in a consistent, innovative stock-keeping behavior.


In the next section, we compare the performance of the three algorithms: tabular Q-Learning (\textit{single\_Q}), centralised A2C (\textit{single\_PG}) and decentralised A2C (\textit{multi\_PG}) in terms of their convergence and execution time.

\subsubsection{Convergence}
Convergence analysis shows the speed at which agents learn to maximise rewards. In other words, how many training episodes are needed before agents learn the optimum behavior. In this experiment, the number of time steps per episode is kept at 1000. 

\begin{figure} 
\centering    
\includegraphics[width=0.4\textwidth]{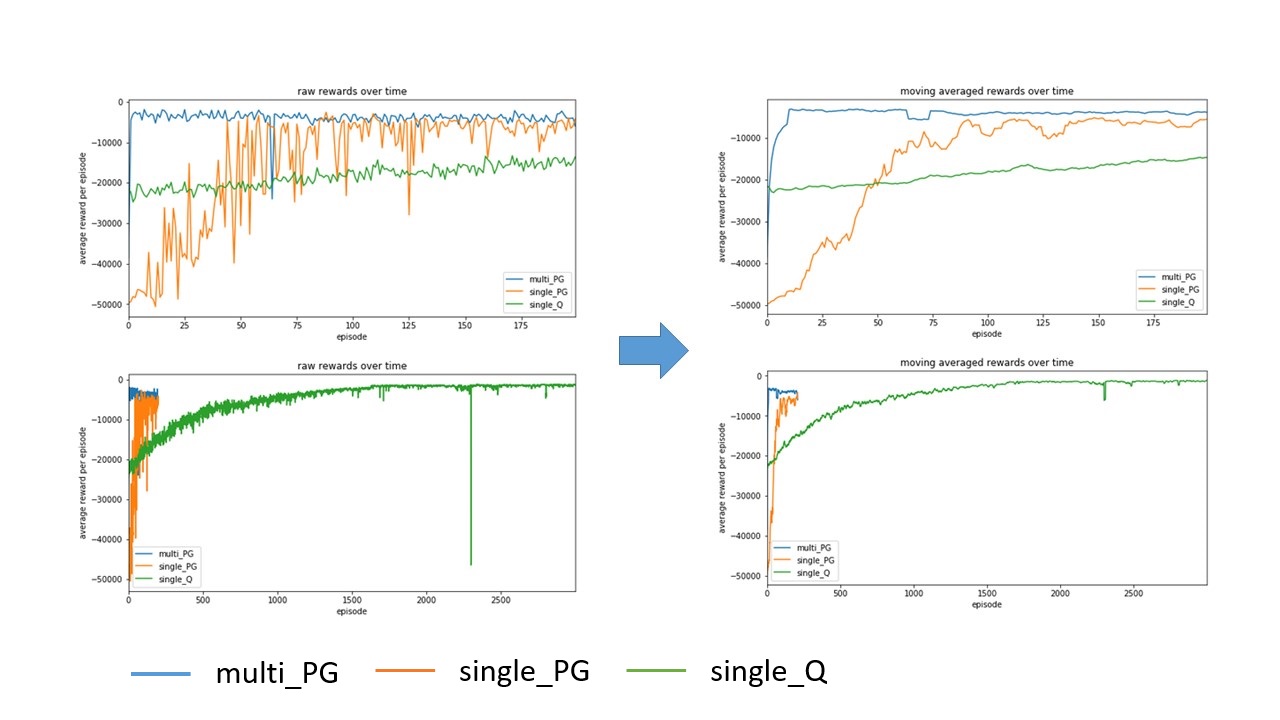}
\caption*{(a) Convergence Comparison Case 1. (Left) Raw rewards over iterations until 200 and 3000 episodes. (Right) Moving Average of 10 rewards over iterations obtained from the left charts.}

\bigskip

\includegraphics[width=0.4\textwidth]{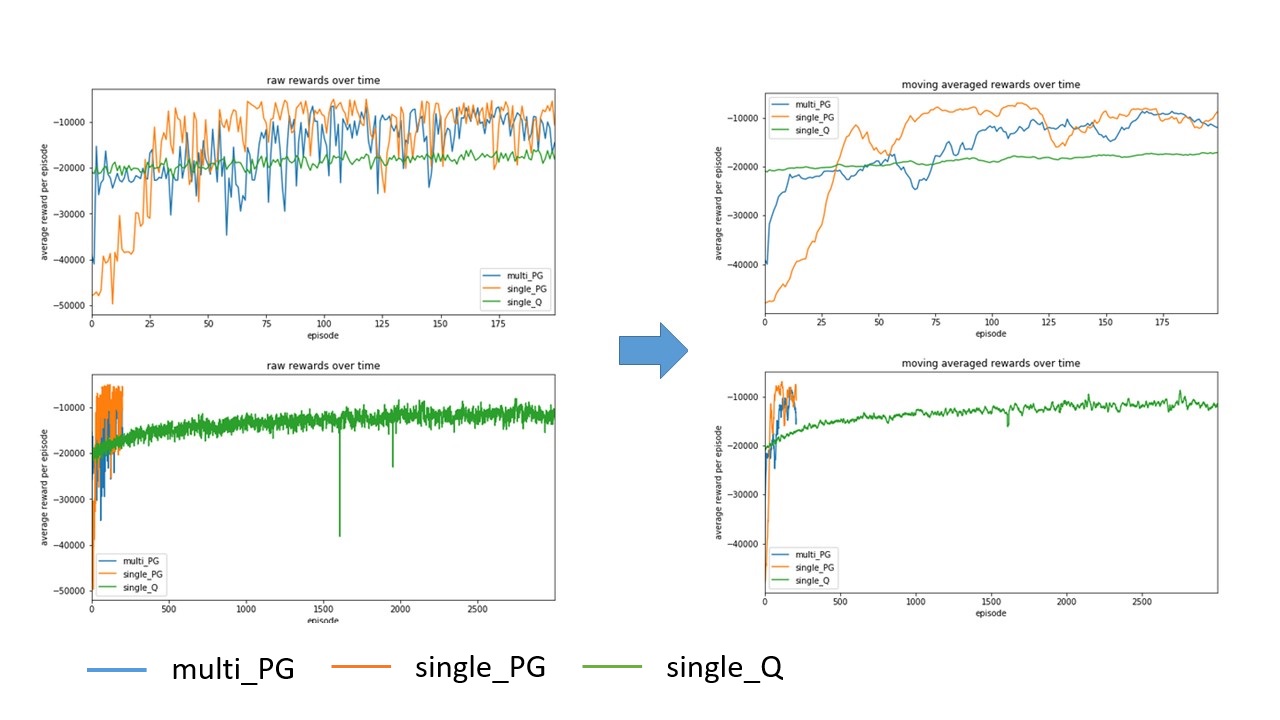}
\caption*{(b) Convergence Comparison Case 2. (Left) Raw rewards over iterations until 200 and 3000 episodes. (Right) Moving Average of 10 rewards over iterations obtained from the left charts.}

\bigskip

\caption{Convergence Comparison}
\label{fig:comparison-convergence}
\end{figure}

Figure \ref{fig:comparison-convergence}(a) and \ref{fig:comparison-convergence}(b) show that both centralised and decentralised A2C methods require less episodes than tabular Q-Learning to converge. After about 200 episodes, the former has converged while the latter needs about 3000 episodes to reach the same behavior.

\subsubsection{Execution Time}

\begin{figure}[htbp!] 
\centering    
\includegraphics[width=0.5\textwidth]{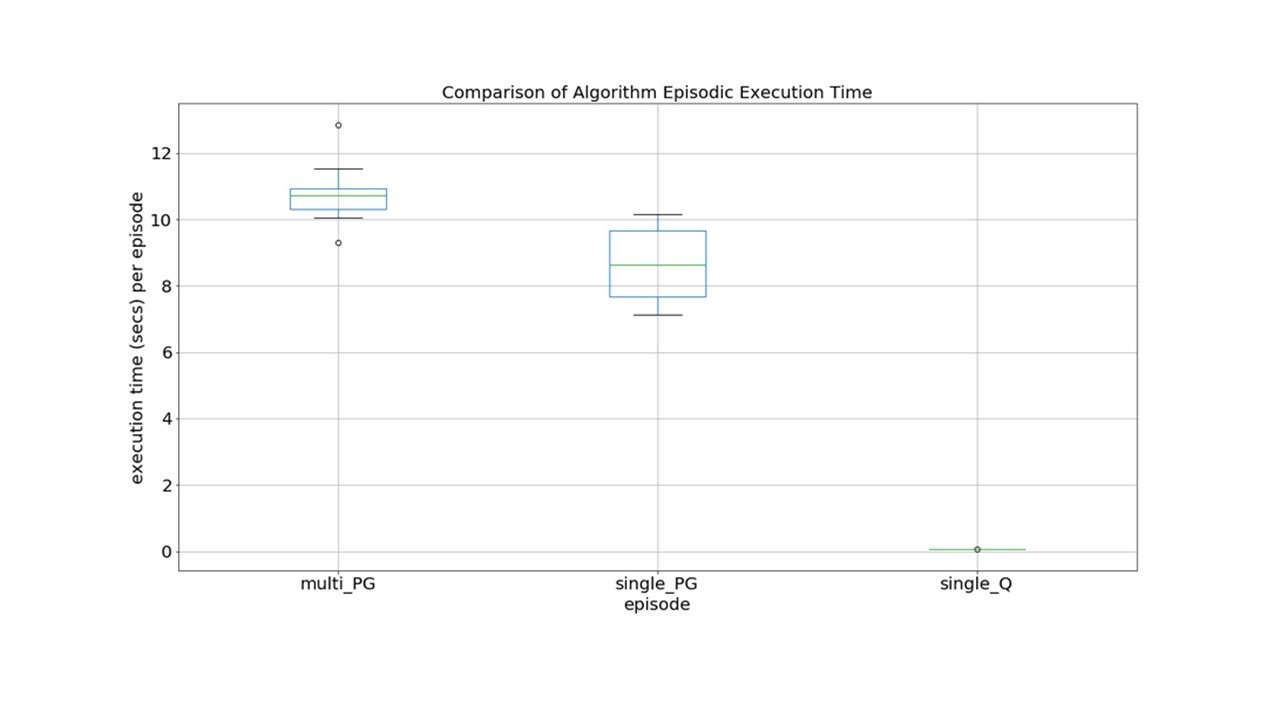}
\caption{Execution Time Comparison }
\label{fig:comparison-executiontime}
\end{figure}

Execution time refers to the amount of learning time that each algorithm requires for every episode. For tabular Q-learning this refers to the time it needs to update each table entry. Meanwhile for A2C this is the time taken to perform backpropagation. Figure \ref{fig:comparison-executiontime} shows that Tabular Q-Learning is faster than both centralised and decentralised A2C. This is expected since tabular Q-Learning performs a simple table lookup and update. Note that in this experiment we only use CPU for training. Due to the nature of backpropagation, the execution time of both A2C methods can be significantly improved if GPU parallelisation is used.

Based on the analysis done in this chapter, we can conclude that Multi-Agent A2C uses less information than both A2C and Tabular Q-Learning. In particular, the former only requires each agent to see their local inventory level and customer demand, whereas the latter need to see every other agents' inventory level. In terms of convergence, both Multi-Agent A2C and A2C require less episodes to converge, since these algorithms generalise better than Tabular Q-Learning. Nonetheless, when it comes to actual execution time, Tabular Q-Learning still triumphs over the others. More fundamental research is needed to speed up the two A2C algorithms.

\section{Discussion and Conclusions}

The safety stock optimisation problem has been studied for decades. Nevertheless, it is puzzling why many companies, in practice, tend to still use rather simplistic means to calculate safety stock level \cite{csco_five_2011}. One reason is highlighted by \citet{humair_incorporating_2013}, where most emphases have been put on deriving algorithms with provable optimality, often at the expense of simplifying reality. However, companies are typically not as concerned about using the most optimally proven yet oversimplified policy. Instead, they prefer to utilise solutions that accommodates their actual pain points.

We postulate that Reinforcement Learning (RL) algorithms may have a role to play in safety stock optimisation as its strength lies in the types of challenges companies find in adopting optimisation approaches to safety stock problems they face. These include ability to handle complex environments dynamically and as such, being able to to optimise both safety stock level and order quantity rule simultaneously. Such facets of the safety stock problem have not been well-studied in literature. It is often unrealistically assumed that actors forming the various echelons of a supply chain all use the same simple order quantity rule. RL mitigates these assumptions by its nature. 

To measure the efficacy of RL for safety stock optimisation, we designed a problem involving a serial structure with three agents. Three RL algorithms have been adopted to the problem including tabular Q-Learning, TD Advantage Actor Critic (A2C) and multi-agent TD A2C. The agents are trained to maximise their joint rewards i.e. minimising total inventory cost and stock-out levels. The trained RL algorithm results in two neural networks: actor and critic. The former governs the agents' order quantity rule while the latter calculates the desired safety stock level, all conditioned upon the current inventory levels of all agents. The algorithms were validated using a mathematical programming approach as the baseline, which assumed \textit{a priori} that the order quantity rule follows a base stock policy e.g. equivalent to base stock - current inventory. 

However, we have also encountered several challenges with RL-based optimisation algorithms that are worthy for future investigation. First, while neural network based policies could capture more complex behaviour, they are often  difficult to interpret. In the wider literature of machine learning, this is often referred to as the explainability problem. Explainability is a concern for practitioners, as it reduces trust in the algorithm and its output, especially because optimality of an RL-based solution cannot be mathematically proven – it is typically assumed that convergence of reward is equivalent to reaching optimality. Second, the convergence rate and execution time of RL-based models are still slow relative to analytical-based mathematical programming approaches. It is currently still difficult to scale RL to a very large number of agents. However, researchers have actively been working on tackling this problem. Our experiment shows that certain classes of RL, like the Multi-Agent A2C, could scale better than traditional Q-Learning. Nevertheless, more investigation is needed to improve its' execution time. Therefore practically speaking, in the current state, there is a trade off between deploying an RL algorithm for a more realistic, out of the box solutions, and the time it takes to arrive at solutions. 

Given these challenges, there are several extensions to our current approach that could be further explored. These include: experimentation with more complex supply chain structures, incorporating general stochastic demand patterns reflecting Enterprise Resources Planning (ERP) system data, and incorporating more complex customer behaviour such as backordering. 

Furthermore, extension of the model to allow local-based information would be a natural progression. With the Actor-Critic framework, this would translate into using actor and critic that are both decentralised, such that each supply agent pursues their own goals with locally available information.

{\small
	\bibliographystyle{apalike}
	\bibliography{dp.bib}
}


\appendix

\section{Reinforcement Learning Implementations}

Table \ref{table:RL-variables} defines the variables that will be used in the three RL algorithms. The three algorithms: Q-Learning, Temporal Difference (TD) Advantage Actor-Critic (A2C) and Multi-agent TD A2C are written below.

\begin{center}
\begin{table}[ht]
\centering
\caption{Reinforcement Learning Variable Definitions}
\begin{tabular}{ | m{6cm}| m{1cm} | } 
\hline
policy function & $\pi()$ \\
\hline
action-value function & $Q()$ \\ 
\hline
action-value function at a given policy $\pi$ & $Q_{\pi}()$ \\ 
\hline
state-value function & $V()$ \\ 
\hline
state-value function at a given policy $\pi$ & $V_{\pi}()$ \\ 
\hline
time & $t$ \\
\hline
step reward & $r$ \\
\hline
step reward at time t& $r_t$ \\
\hline
Total reward over T steps & $R_T$ \\
\hline
action & $a$ \\
\hline
action at time t& $a_t$ \\
\hline
state & $s$ \\ 
\hline
state at time t& $s_t$ \\
\hline
learning rate & $\alpha$ \\
\hline
discount factor & $\gamma$ \\
\hline
greedy noise factor & $\epsilon$ \\
\hline
neural network weights & $W$ \\
\hline
policy (neural) network weights & $\theta$ \\
\hline
value (neural) network weights & $w$ \\
\hline
\end{tabular}
\label{table:RL-variables}
\end{table}
\end{center}

\vspace{3ex}
\begin{algorithm}
\SetAlgoLined
\KwResult{$Q(s, a)$ for $s \in States$, $a \in Actions(s)$ }
 Hyperparameters: learning rate $\alpha \in (0, 1]$, discount factor $\gamma \in (0, 1]$, epsilon-greedy factor $\epsilon \in (0, 1]$ \;
 Initialise $Q(s, a)$ for $s \in States$, $a \in Actions(s)$ arbitrarily \;
 \For{each episode}{
  initialise $s_{factory}$\;
  \For{ $t$ in $[0, T]$}{
      \eIf{$random.random() < \epsilon$}{
          $a_t$ = randomly selected from Actions($s_t$)\;
      }{
          $a_t$ = $\underset{a}{\arg\max}(Q(s_t, a))$\;
      }
      take action $a_t$, observe $r_{t+1}$ and $s_{t+1}$ \;
      find off-policy action $a'_{t+1} = \underset{a}{\arg\max} (Q(s_{t+1}, a))$ \;
      $Q(s_t, a_t) \leftarrow Q(s_t, a_t) + \alpha [r_{t+1} + \gamma Q(s_{t+1}, a'_{t+1}) - Q(s_t, a_t)]$ \;
  }
 }
 \caption{Q-Learning}
\label{algo:q-learning}
\end{algorithm}
\vspace{3ex}

\vspace{3ex}
\begin{algorithm}
\SetAlgoLined
\KwResult{$V(s)$ and $\pi(a|s)$ for $s \in States$ and $a \in Actions(s)$  }
 Hyperparameters: learning rate $\alpha^\theta, \alpha^w \in (0, 1]$, discount factor $\gamma \in (0, 1]$, standard deviation $\sigma$ \;
 Initialise policy network weights $\theta$ and value network weights $w$ arbitrarily \;
 \For{each episode}{
  initialise $s_{factory}$\;
  \For{ $t$ in $[0, T]$}{
      $a_t \sim \pi(\cdot|s_t, \theta)$ \;
      take action $a_t$, observe $r_{t+1}$ and $s_{t+1}$ \;
      $\delta = r_{t+1} + \gamma V(s_{t+1}|w) - V(s_t|w)$ \;
      update value parameter $w \leftarrow w + \alpha^w \delta \nabla V(s_t, w)$ \;
      update policy parameter $\theta \leftarrow \theta + \alpha^\theta \delta \nabla \ln{\pi(a_t|s_t, \theta)}$ \;
  }
 }
 \caption{TD Advantage Actor-Critic (A2C)}
\label{algo:centralised-a2c}
\end{algorithm}
\vspace{3ex}

\vspace{3ex}
\begin{algorithm}
\SetAlgoLined
\KwResult{$V([s^1, \dots, s^N])$ and $\pi^i(a^i|s^i)$ for $s^i \in States$ and $a^i \in Actions(s^i)$ for all agents $i \in \{1, \dots, N\}$  }
 Hyperparameters: learning rate $\alpha^\theta = \alpha^{\theta^1} = \dots = \alpha^{\theta^N}, \alpha^w \in (0, 1]$, discount factor $\gamma \in (0, 1]$, standard deviation $\sigma$ \;
 Initialise policy network weights ${\theta^1, \dots, \theta^i, \dots, \theta^N} $ for each agent $i \in {1, \dots, N}$ and value network weights $w$ arbitrarily \;
 \For{each episode}{
  initialise $S_{factory} = [s^1_0, \dots, s^N_0]$\;
  \For{ $t$ in $[0, T]$}{
      \For{each agent $i$ in $\{1, \dots, N\}$}{
        $a^i_t \sim \pi^i(\cdot|s^i_t, \theta^i)$ \;
      }
      take action $a_t = [a^1_t, \dots, a^N_t]$, observe joint reward $r_{t+1}$ and $s_{t+1} = [s^1_t, \dots, s^N_t]$\;
      joint $\delta = r_{t+1} + \gamma V(s_{t+1}|w) - V(s_t|w)$ \;
      update value parameter $w \leftarrow w + \alpha^w \delta \nabla V(s_t, w)$ \;
      \For{each agent $i$ in $\{1, \dots, N\}$}{
        update policy parameter $\theta^i \leftarrow \theta^i + \alpha^\theta \delta \nabla \ln{\pi^i(a^i_t|s^i_t, \theta^i)}$ \;
      }

  }
 }
 \caption{Multi-Agent TD Advantage Actor-Critic (A2C)}
\label{algo:decentralised-a2c}
\end{algorithm}
\vspace{3ex}

\end{document}